# Who Will Retweet This? Automatically Identifying and Engaging Strangers on Twitter to Spread Information

**Kyumin Lee**
Department of Computer Science
Utah State University
Logan, UT 84322
kyumin.lee@usu.edu

**Jalal Mahmud, Jilin Chen, Michelle Zhou, Jeffrey Nichols**
IBM Research – Almaden
San Jose, CA 95120
{jumahmud, jilinc, mzhou, jwnichols}@@us.ibm.com

**ABSTRACT**

There has been much effort on studying how social media sites, such as Twitter, help propagate information in different situations, including spreading alerts and SOS messages in an emergency. However, existing work has not addressed how to *actively* identify and engage the right strangers at the right time on social media to help effectively propagate intended information within a desired time frame. To address this problem, we have developed two models: (i) a feature-based model that leverages peoples' exhibited social behavior, including the content of their tweets and social interactions, to characterize their willingness and readiness to propagate information on Twitter via the act of retweeting; and (ii) a wait-time model based on a user's previous retweeting wait times to predict her next retweeting time when asked. Based on these two models, we build a recommender system that predicts the likelihood of a stranger to retweet information when asked, within a specific time window, and recommends the top-*N* qualified strangers to engage with. Our experiments, including live studies in the real world, demonstrate the effectiveness of our work.

**ACM Classification Keywords**
H.5.2. [Information Interfaces and Presentation]: User Interfaces.

**INTRODUCTION**

With the widespread use of social media sites, like Twitter and Facebook, and the ever growing number of users, there has been much effort on understanding and modeling information propagation on social media [1, 2, 6, 14, 17, 25, 27, 29, 30]. Most of the work assumes that information is propagated by a small number of influential volunteers, who possess certain qualities, such as having a large number of followers, which make them extremely effective in propagating information [28]. For example, these users can help spread emergency alerts, such as fire hazard or SOS messages like requesting blood donations, to reach more people faster.

However, prior research efforts ignore several critical factors in influencer-driven information propagation. First, influential users may be unwilling to help propagate the intended information for various reasons. For example, they may not know the truthfulness of a piece of information, and thus are unwilling to risk their reputation to spread the information. Second, an influential user may be unavailable to help propagate information when needed. For example, influential users may not be online to help propagate SOS messages when a disaster strikes.

Since everyone is potentially an influencer on social media and is capable of spreading information [2], our work aims to identify and engage the right people at the right time on social media to help propagate information when needed. We refer to these people as *information propagators*. Since not everyone on social media is willing or ready to help propagate information, our goal is to model the characteristics of information propagators based on their social media behavior. We can then use the established model to predict the likelihood of a person on social media as an information propagator. As the first step, we focus on modeling *domain-independent* traits of information propagators, specifically, their *willingness* and *readiness* to spread information.

In many situations including emergency or disastrous situations, information propagation must be done within a certain time frame to optimize its effect. To satisfy such a time constraint, we thus also develop a wait-time model based on a user's previous retweeting wait times to predict the user's next retweeting time when asked.

For the sake of concreteness, in this paper we focus on Twitter users, although our core technology can be easily applied to other social media platforms. On Twitter, the most common method for propagating information is retweeting[1], which is to repost others' tweets in your own content stream. Our work is thus reduced to the problem of finding strangers on Twitter who will retweet a message when asked.

To model one's willingness and readiness to retweet information, we first identify a rich set of features to characterize the candidate, including derived personality traits, social network information, social media activity, and previous retweeting behavior. Unlike existing work, which often uses only social network properties, our feature set includes *personality traits* that may influence one's retweeting behavior. For example, when asked by a stranger in an emergency, a person with a high level of altruism may be more responsive and willing to retweet. Similarly, a more active user who frequently posts status updates or reposts others' tweets may be more likely to retweet when asked. Our features capture a variety of characteristics that are likely to influence one's retweeting behavior.

To predict one's likelihood to retweet when asked, we train statistical models to infer the weights of each feature, which are then used to predict one's likelihood to retweet. Based on the prediction models, we also build a real-time recommender system that can rank and recommend the top-$N$ candidates (*retweeters*) to engage with on Twitter.

To demonstrate the effectiveness of our work, we have conducted extensive experiments, including live studies in the real world. Compared to two baselines, our approach significantly improves the *retweeting rate*[2]: the ratio between the number of people who retweeted and the number of people asked. To the best of our knowledge, our work is the first to address how to *actively* identify and engage strangers on Twitter to help retweet information. As a result, our work offers three unique contributions:

- A feature-based model including one's personality traits for predicting the likelihood of a stranger on Twitter to retweet a particular message when asked.
- A wait-time model based on a person's previous retweeting wait times to estimate her next retweeting wait time when asked.
- A retweeter recommender system that uses the two models mentioned above to effectively select the right set of strangers on Twitter to engage with in real time.

**RELATED WORK**

Our work is most closely related to the recent efforts on actively engaging strangers on social media for accomplishing certain tasks [22, 23]. However, ours is the first on modeling and engaging strangers on social media to aid information propagation within a given time window.

Our work is also related to the effort on characterizing retweeters and their retweeting behavior [21]. However, the existing work does not include personality features as our model does. More importantly, unlike the existing model focusing on *voluntary* retweeting behavior, ours examines a person's retweeting behavior at the request of a stranger.

There are many efforts on modeling influential behavior in social media. Such work finds influential users by their social network properties [2, 6, 14, 17, 27, 30], content of posts [1], information forwarding/propagating activity [25], and information flow [29]. In comparison, our work focuses on an individual's characteristics that influence their willingness and readiness to retweet at a stranger's request. Some of these characteristics, such as personality and readiness to retweet, have not been studied before.

As our goal is to support effective information diffusion, our work is related to efforts in this space. Bakshy et al. [3] examine the role of the social network and the effects of tie strength in information diffusion. Chaoji et al. [7] show how to maximize content propagation in one's own social network. In contrast, our approach aims at selecting a right set of *strangers* on social media to help spread information. Budak et al. [5] have studied a different type of information diffusion, which spreads messages to counter malicious influences, and hence minimize the influence of such campaigns. They proposed to identify a subset of individuals to start a counter campaign based on a set of viral diffusion features, including user virality and susceptibility, and item virality [16]. These features are complementary to the features that we use, such as personality, messaging activity, and past retweeting activity. Moreover, there is little work on automatically identifying and engaging the *right* strangers at the *right* time on social media to aid information propagation as ours does.

**CREATING GROUND-TRUTH DATASETS**

Since there is no publicly available ground-truth data with which we can train and build our predictive models, we collected two real-world datasets. We created a total of 17 Twitter accounts and our system automatically sent retweeting requests to 3,761 strangers on Twitter. Our first data set examines *location-based targeting*, where people who live in a particular location were asked to retweet information relevant to that location. The second examines *topic-based targeting*, where people interested in a certain topic were asked to retweet information relevant to that topic.

We hypothesize that information relevance influences a person's retweeting behavior especially at the request of a stranger. For example, people might be more likely to retweet news about public safety in an area where they live or work rather than for other locations. Similarly, a person might be more willing to retweet information on a topic in which s/he is interested.

Our dataset for location-based targeting (named *"public safety"*) and the dataset for topic-based targeting (named *"bird flu"*) are intended to examine how different types of information (location vs. topic) may impact retweeting behavior.

---

[1] We use the term "repost", "retweet" and "propagate" interchangeably

[2] We use the term "information propagation rate", "information repost rate" and "retweeting rate" interchangeably

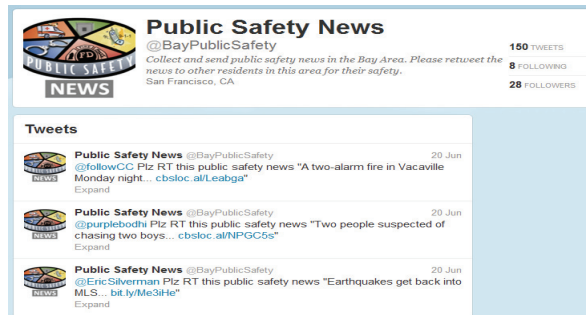

Figure 1. An example Twitter account created for Public Safety data collection.

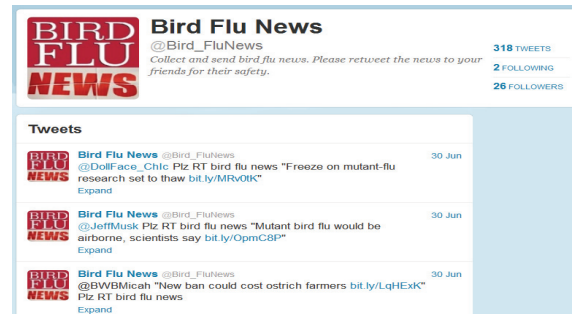

Figure 2. An example Twitter account created for Bird Flu data collection.

**Public Safety Data Collection**: For location-based targeting, we chose the San Francisco bay area as the location and sent tweets about local public safety news to people whom we identified as living or staying in that area. First, we created 9 accounts on Twitter. All accounts had the same profile name, "Public Safety News" and the same description (Figure 1).

Note that we created multiple accounts to send a few messages per hour from each account in order to create a reasonable pretense of human behavior. Furthermore, previous studies have shown that if not careful, target strangers would silently flag an account as a spam to cause the suspension of the account by Twitter [22, 23]. Creating multiple accounts helped us avoid this possibility, and thus increased the number of users that we could reasonably contact per hour (each user received only one message).

Creating multiple accounts for research purposes is a commonly used methodology [18, 19]. To make these accounts appear to be genuine, all accounts followed 4~10 users and had 19 followers. We also created the following and follower accounts, and some were also followed by the original accounts. We posted 11 public safety messages using each of the 9 accounts before we contacted anyone on Twitter. We identified 34,920 bay area Twitter users using the Twitter Streaming API[3] with a geo-location filter corresponding to the bay area in June 2012. This stream retrieved only tweets that were marked as being sent within a bounding box equivalent to the bay area determined by using the Google Geocoding API[4]. We filtered out non-English tweets in this stream, and created a list of unique users whose tweets were in the stream.

Among all the identified Twitter users, we randomly selected 1,902 people. From our public safety accounts, our system automatically sent messages to those people using the Twitter API and ensured that each person received only one message to avoid overburdening the person. Here is an example message sent:

@ SFtargetuser "A man was killed and three others were wounded in a shooting... http://bit.ly/KOl2sC" Plz RT this safety news

Each message contained the target person's screen name, the title of a news article obtained from a local news media site, a link to the article, and a phrase asking the person to retweet the message. The original link URL was shortened with the bit.ly URL shortening service to allow us to track user clicks on the link. Per our requests, 52 of the 1,902 (2.8%) people retweeted our message, which reached a total of 18,670 followers of theirs.

**Bird Flu Data Collection**: for topic-based targeting, we chose people who tweeted about "bird flu", a topic commonly being discussed at the time of our study. First, we created 8 accounts on Twitter (Figure 2). All accounts followed 2~5 users and had 19 followers. The following and followers accounts were created using the same method as in the public safety scenario. We then collected 13,110 people's profiles using the Twitter Search API and the queries "bird flu", "H5N1" and "avian influenza" in June 2012. We excluded non-English tweets and randomly selected 1,859 users. A message was then automatically sent to each selected person. Here is an example message sent:

@birdflutargetuser Plz RT bird flu news "Bird Flu viruses could evolve in nature http://bit.ly/MQBASY"

As in the public safety study, the news articles were obtained from the news media sites. 155 of the 1,859 users (8.4%) retweeted our messages, which reached their 184,325 followers.

For both datasets, through the Twitter API we collected publicly available information of each person whom we asked to retweet. This included their profile, people they followed, followers, up to 200 of their most recently posted messages, and whether they retweeted our message (the ground truth).

## FEATURE EXTRACTION

To model a person's likelihood to retweet, we have identified six categories of features, as described below.

---

[3] http://dev.twitter.com/pages/streaming_api

[4] https://developers.google.com/maps/documentation/geocoding/

**Profile Features**
Profile features are extracted from a user's Twitter profile and consist of: *longevity (age) of an account*, *length of screen name*, *whether the user profile has a description*, *length of the description*, and *whether the user profile has a URL*. Our hypothesis behind the use of these features is that a user with a richer profile or a longer account history may be more knowledgeable in using advanced social media features, such as retweeting. Hence, when asked, they are more likely to retweet than those who have just opened an account recently or have little information in their profile.

**Social Network Features**
We use the following features to characterize a user's social network: *number of users following (friends)*, *number of followers*, and the *ratio of number of friends to number of followers*. These features indicate the "socialness" of a person. Intuitively, the more social a person is (e.g., a good number of followers), the more likely the person may be willing to retweet. These features may also signal potential motivations for retweeting (e.g., an act of friendship and to gain followers) [4]. However, a person (e.g., a celebrity) with an extraordinary number of followers may be unwilling to retweet per a stranger's request.

**Personality Features**
Researchers have found that word usage in one's writings, such as blogs and essays, are related to one's personality [11, 13, 24]. Using the approach described in [22], we computed 103 personality features from one's tweets: 68 LIWC features (e.g., word categories such as "sadness") [24], and 5 Big5 dimensions (e.g., *agreeableness* and *conscientiousness*) with their 30 sub-dimensions [10, 32]. These features may signal potential motivations for retweeting (e.g., an act of altruism and to gain followers) [4].

**Activity Features**
This feature category captures people's social activities. Similar to the reasons stated earlier, our hypothesis is that the more active people are, the more likely they would retweet when asked by a stranger. Moreover, new Twitter users or those who rarely tweet may not be familiar with the retweeting feature and be less likely to reweet. To evaluate this hypothesis, we use the following features:

- *Number of status messages*
- *Number of direct mentions (e.g., @johny) per status message*
- *Number of URLs per status message*
- *Number of hashtags per status message*
- *Number of status messages per day during her entire account life (= total number of posted status messages / longevity)*
- *Number of status messages per day during last one month*
- *Number of direct mentions per day during last one month*
- *Number of URLs per day during last one month*
- *Number of hashtags per day during last one month*

These features also help us distinguish "sporadic" vs. "steady" activeness. We hypothesize that "steady" users are more dependable and are more likely to retweet when asked. For each person, we computed these features based on their 200 most recent tweets, as our experiments have shown that 200 tweets are a good representative sample for deriving one's features.

**Past Retweeting Features**
We capture retweeting behavior with the these features:

- *Number of retweets per status message: R/N*
- *Average number of retweets per day*
- *Fraction of retweets for which original messages are posted by strangers who are not in her social network*

Here $R$ is the total number of retweets and $N$ is the total number of status messages. We hypothesize that frequent retweeters are more likely to retweet in the future.

The third feature measures how often a person retweets a message originated outside of the person's social network. We hypothesize that people who have done so are more likely to retweet per a stranger's request to do so.

**Readiness Features**
Even if a person is willing to retweet per a request, he may not be ready to do so at the time of the request due to various reasons, such as being busy or not being connected to the Internet. Since such a context could be quite diverse, it is difficult to model one's readiness precisely. We thus use the following features to approximate readiness based on one's previous activity:

- *Tweeting Likelihood of the Day*
- *Tweeting Likelihood of the Hour*
- *Tweeting Likelihood of the Day (Entropy)*
- *Tweeting Likelihood of the Hour (Entropy)*
- *Tweeting Steadiness*
- *Tweeting Inactivity*

The first two features are computed as the ratio of the number of tweets sent by the person on a given day/hour and the total number of tweets. The third and fourth features measure entropy of tweeting likelihood of the day and the hour, respectively [26]. Below is a person's ($u$) entropy of tweeting likelihood of the hour $P(x_1), P(x_2), P(x_3) ... P(x_n)$:

$$Entropy(u) = -\sum_{i=1}^{n} P(x_i) \log P(x_i)$$

In the above equation, $n$ is 24 to estimate the daily likelihood to tweet. The tweeting steadiness feature is computed as $1/\sigma$, where $\sigma$ is the standard deviation of the elapsed time between consecutive tweets, computed from the most recent $K$ tweets (where $K$ is set to 20). The tweeting inactivity feature is the difference between the time when a retweeting request is sent and the time when user last tweeted.

## PREDICTING RETWEETERS

Based on the features described above, we train a model to predict a user's likelihood to be a retweeter.

**Training and Test Set.** First we randomly split each dataset (public safety and bird flu) into training (containing 2/3 data) and testing sets (containing 1/3 data). The two sets were stratified, and contained the same ratio of retweeters and non-retweeters. Finally for public safety, the training set had 35 retweeters and 1,233 non-retweeters; and the test set had 17 retweeters and 617 non-retweeters. For bird flu, the training set had 103 retweeters and 1136 non-retweeters; the test data had 52 retweeters and 568 non-retweeters. For each person in the sets, we computed all the features described previously.

**Predictive Models.** We compared the performance of five popular models: Random Forest, Naïve Bayes, Logistic Regression, SMO (SVM), and AdaboostM1 (with random forest as the base learner). We used WEKA [15] implementation of these algorithms and trained these models to predict the probability of a person to retweet and classify a person as a retweeter or non-retweeter.

**Handling Class Imbalance**. Both our datasets have an imbalanced class distribution: only 52 out of 1,902 users (2.8%) in the public safety dataset and 155 out of 1,859 users (8.4%) in the bird flu dataset were retweeters. Imbalanced class distribution in a training set hinders the learning of representative sample instances, especially the minority class instances, and prevents a model from correctly predicting an instance label in a testing set. The class imbalance problem has appeared in a large number of domains, such as medical diagnosis and fraud detection. There are several approaches to the problem, including over-sampling minority class instances, under-sampling majority class instances, and adjusting the weights of instances. Currently, we used both over-sampling and weighting approaches to our class imbalance problem. For over-sampling, we used the SMOTE [8] algorithm. For weighting, we used a cost-sensitive approach of adding more weight to the minority class instances [20].

**Feature Analysis.** To improve the performance of our models, we analyzed the significance of our features using the training set. We computed the $\chi^2$ value for each feature to determine its discriminative power [31], and eliminated the features that do not contribute significantly to the result. Our analyses found 21 and 46 significant features for the two data sets, respectively (Tables 1 and 2). Moreover, several feature groups have more significant power distinguishing between retweeters and non-retweeters: *activity*, *personality*, *readiness*, and *past retweeting*. Although our two datasets are quite different, we found six significant features common to both sets (bolded in Tables 1 and 2). This suggests that it is possible to build *domain-independent* models to predict retweeters. In addition, our analysis suggests that retweeters are more advanced Twitter users, since they use advanced features more frequently (e.g., inclusion of URLs and hashtags in their tweets).

| Feature Group | Significant Features (bolded is common to both data sets) |
|---|---|
| Profile | the longevity of the account |
| Social-network | \|following\| |
| | ratio of number of friends to number of followers |
| Activity | **\|URLs\| per day** |
| | **\|direct mentions\| per day** |
| | **\|hashtags\| per day** |
| | \|status messages\| |
| | \|status messages\| per day during entire account life |
| | \|status messages\| per day during last one month |
| Past Retweeting | **\|retweets\| per status message** |
| | **\|retweets\| per day** |
| Readiness | Tweeting Likelihood of the Day |
| | Tweeting Likelihood of the Day (Entropy) |
| Personality | 7 LIWC features: **Inclusive**, Achievement, Humans, Time, Sadness, Articles, Nonfluencies |
| | 1 Facet feature: Modesty |

**Table 1. 21 Features Selected by $\chi^2$ in Public Safety Dataset**

| Feature Group | Significant Features (bolded is common to both data sets) |
|---|---|
| Profile | the length of description |
| | has description in profile |
| Activity | **\|URLs\| per day** |
| | **\|direct mentions\| per day** |
| | **\|hashtags\| per day** |
| | \|URLs\| per status message |
| | \|direct mentions\| per status message |
| | \|hashtags\| per status message |
| Past Retweeting | **\|retweets\| per status message** |
| | **\|retweets\| per day** |
| | \|URLs\| per retweet message |
| Readiness | Tweeting Likelihood of the Hour (Entropy) |
| Personality | 34 LIWC features: **Inclusive**, Total Pronouns, 1st Person Plural, 2nd Person, 3rd Person, Social Processes, Positive Emotions, Numbers, Other References, Occupation, Affect, School, Anxiety, Hearing, Certainty, Sensory Processes, Death, Body States, Positive Feelings, Leisure, Optimism, Negation, Physical States, Communication |
| | 8 Facet features: Liberalism, Assertiveness, Achievement Striving, Self-Discipline, Gregariousness, Cheerfulness, Activity Level, Intellect |
| | 2 Big5 features: Conscientiousness, Openness |

**Table 2. 46 Features Selected by $\chi^2$ in Bird Flu Dataset**

## Incorporating Time Constraints

While our predictive models compute a person's likelihood to retweet upon request, it does not predict when that person will retweet. Some situations may require important messages to be spread quickly, such as emergency alerts and SOS messages, so we also explore how to predict when a person will act on the retweeting request. To do this, we examine the person's previous temporal behavior and use this information for prediction.

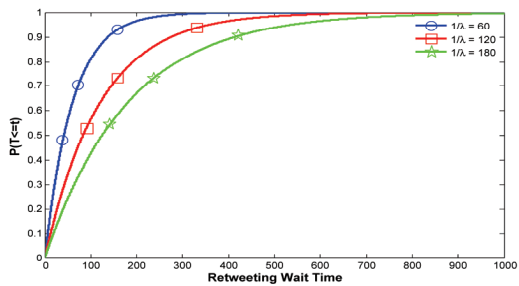

Figure 3. Three examples of the exponential distribution.

In the simplest case, our model estimates the wait time for a person to respond to a retweeting request. We further assume that retweeting events follow a poisson process during which each retweeting occurs continuously and independently at a constant average rate. We thus use an exponential distribution model to estimate a user's retweeting wait time with a probability. The cumulative distribution function (CDF) of an exponential distribution is:

$$f(x;\lambda) = \begin{cases} 1 - e^{-\lambda x}, & x \geq 0, \\ 0, & x < 0. \end{cases}$$

The distribution is on the interval from zero to infinite. We measure $\frac{1}{\lambda}$ which is the average wait time for a user based on prior retweeting wait time. For a user's specific retweeting wait time $t$, our model can predict the probability of the user's next retweeting $P(t)$ within that wait time. Figure 3 shows our model with three examples. The green line with stars indicates that a person's average wait time is 180 minutes based on past retweeting behavior. The retweeting probability within 200 minutes is larger than 0.6. The lower a person's average retweeting wait time $t$ is, the higher probability of her retweeting is within time $t$.

In practice, given a specific time constraint $t$, we select a *cut-off probability c* that is then used to select people whose probability of retweeting within time $t$ is greater than or equal to $c$. For example, with the cut-off probability of 0.7, our model will select only those who have at least 70% chance to retweet within the given time constraint. Incorporating the time estimation with our prediction models, we contact only people who are likely to retweet and whose cumulative probability of the retweeting wait time is greater than or equal to the *cut-off probability c*.

### Incorporating Benefit and Cost

We have also explored the trade-offs between the cost of contacting a user and the benefit of a re-tweet. We assume the benefit is the number of people who are directly exposed to the message as a result of the re-tweets, which is the total number of followers of the retweeter. Using this assumption, if our system contacts $N$ users and $K$ retweet, the total benefit is then the sum of all followers of the $K$ users. Assuming a unit cost per contact, the total cost is then $N$. We normalize the total benefit by total cost to compute *unit-info-reach-per-person*:

$$\text{unit-info-reach-per-person} = \frac{\sum_{1}^{K} followers(i)}{N}$$

To address the case that the same person follows multiple retweeters, we count just the number of *distinct* followers for each retweeter.

### REAL-TIME RETWEETER RECOMMENDATION

As mentioned earlier, our goal is to automatically identify and engage the right strangers at the right time on social media to help spread intended messages within a given time window. We thus have developed an interactive recommender system that uses our prediction model and the wait-time estimation model in *real time* to recommend the right candidates to whom retweeting requests will be sent. Figure 4 shows the interface of our system. Our system monitors the Twitter live stream and identifies a set of candidates who have posted content relevant to the topic of a retweet request (e.g., "bird flu" alerts). Such content filtering can be done by using the approaches detailed in [9]. Based on the identified candidates, our system uses the prediction model to compute the candidates' likelihood of retweeting and their probability of retweeting within the given time window $t$. It then recommends the top-$N$ ranked candidates whose probability of retweeting within $t$ is also greater than or equal to the cut-off probability $c$ (Figure 4a). A user (e.g., an emergency worker) of our system can interactively examine and select the recommended candidates, and control the engagement process, including editing and sending the retweeting request (Figure 4b).

### EXPERIMENTS

We designed and conducted an extensive set of experiments to measure the performance of various prediction models. We also compared the effectiveness of our approach with two base lines in various conditions including a live setting.

### Evaluating Retweeter Prediction

To evaluate the performance of our prediction models, we used only the significant features found by our feature analysis (Tables 1-2) in our experiments.

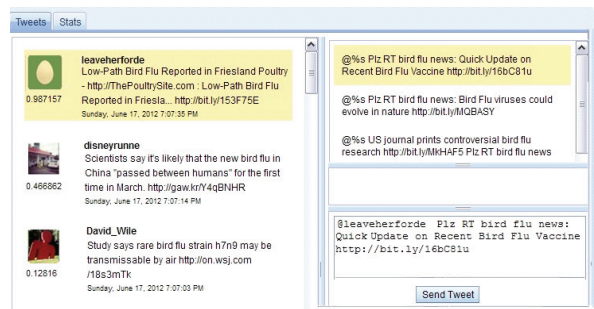

Figure 4. The interface of our retweeter recommendation system: (a) left panel: system-recommended candidates, and (b) right panel: a user can edit and compose a retweeting request.

| Classifier | AUC | F1 | F1 of Retweeter |
|---|---|---|---|
| Basic | | | |
| Random Forest | 0.638 | 0.958 | 0 |
| Naïve Bayes | 0.619 | 0.939 | 0.172 |
| Logistic | 0.640 | 0.958 | 0 |
| SMO | 0.500 | 0.96 | 0 |
| AdaBoostM1 | 0.548 | 0.962 | 0.1 |
| SMOTE | | | |
| Random Forest | 0.606 | 0.916 | 0.119 |
| Naïve Bayes | 0.637 | 0.923 | 0.132 |
| Logistic | 0.664 | 0.833 | 0.091 |
| SMO | 0.626 | 0.813 | 0.091 |
| AdaBoostM1 | 0.633 | 0.933 | 0.129 |
| Cost-Sensitive (Weighting, showing the best results in each model) | | | |
| Random Forest | **0.692** | 0.954 | 0.125 |
| Naïve Bayes | 0.619 | 0.93 | 0.147 |
| Logistic | 0.623 | 0.938 | 0.042 |
| SMO | 0.633 | 0.892 | 0.123 |
| AdaBoostM1 | 0.678 | 0.956 | 0.133 |

Table 3. Prediction accuracy (Public Safety).

| Classifier | AUC | F1 | F1 of Retweeter |
|---|---|---|---|
| Basic | | | |
| Random Forest | 0.707 | 0.877 | 0.066 |
| Naïve Bayes | 0.670 | 0.834 | 0.222 |
| Logistic | 0.751 | 0.878 | 0.067 |
| SMO | 0.500 | 0.876 | 0 |
| AdaBoostM1 | 0.627 | 0.878 | 0.067 |
| SMOTE | | | |
| Random Forest | 0.707 | 0.819 | 0.236 |
| Naïve Bayes | 0.679 | 0.724 | 0.231 |
| Logistic | 0.76 | 0.733 | 0.258 |
| SMO | 0.729 | 0.712 | 0.278 |
| AdaBoostM1 | 0.709 | 0.837 | 0.292 |
| Cost-Sensitive (Weighting, showing the best results in each model) | | | |
| Random Forest | **0.785** | 0.815 | 0.296 |
| Naïve Bayes | 0.670 | 0.767 | 0.24 |
| Logistic | 0.735 | 0.742 | 0.243 |
| SMO | 0.676 | 0.738 | 0.256 |
| AdaBoostM1 | 0.669 | 0.87 | 0.031 |

Table 4. Prediction accuracy (Bird Flu).

**Accuracy Metrics.** We use three metrics to assess prediction accuracy: Area under the ROC Curve (AUC), F1, and F1 of the retweeter class. We use AUC as our primary performance measure, since a higher AUC means that a model is good at correctly predicting both class instances regardless of class imbalance [12]. We report an overall F1 score as a reference measure, and F1 of the retweeter class on the performance of predicting minority class instances.

**Settings.** We ran all five prediction models under three settings: basic, SMOTE, and cost-sensitive.

The *basic* setting did not handle class imbalance. SMOTE was an over-sampling approach in which we over-sampled the minority class instances in the training set such that there was an equal number of majority and minority class instances. Under the *cost-sensitive* setting, we used a weighting scheme that weighted the minority class instances higher than the majority class instances. In our experiments, we tried five different weight ratios from 10:1 through 50:1 at intervals of 10. With five prediction models under three settings, we ran a total of 35 experiments: 5 in the basic setting, 5 in the SMOTE setting, and 25 using the cost-sensitive setting (5 models by 5 weight ratios).

**Prediction Results.** Table 3 shows the results for the public safety dataset. Overall, the cost-sensitive setting (weighting) yielded better performance than SMOTE for both AUC and F1 of the retweeter class. Both random forest and AdaBoostM1 performed particularly well under the cost-sensitive setting. We found the similar results using the bird flu dataset (Table 4). The class imbalance problem can be observed in the poor results under the basic setting. For example, SMO completely failed to predict retweeter instances (F1 of retweeter is 0). Although both SMOTE and the cost-sensitive settings outperformed the basic one, we did not observe any clear advantage of one over the other.

In summary, we have found prediction configurations that produced good results by the measures of AUC and F1. Since Random Forest in the cost-sensitive setting performed the best, we used it in the rest of our experiments.

**Comparison with Two Baselines**

To validate how well our prediction approach helps improve retweeting rate in practice, we compared the retweeting rates produced by our approach with those of two baselines: random people contact and popular people contact.

The *random people contact* approach randomly selects and asks a sub-set of qualified candidates on Twitter (e.g., people living in San Francisco or tweeted about bird flu) to retweet a message. This is precisely the approach that we used during our data collection to obtain the retweeting rates for both data sets. The *popular people contact* approach first sorts candidates in our test set by their follower count in the descending order. It then selects and contacts "popular" candidates whose follower count is greater than a threshold. In our experiment, we chose 100 as the threshold since a recent study reported that more than 87% of Twitter users have less than 100 followers[5]. We also considered other threshold values (e.g., 50, 500, 1000) and found that their retweeting rates were comparable.

---

[5] http://www.beevolve.com/twitter-statistics/

| Approach | Retweeting Rate in Testing Set | |
|---|---|---|
| | Public Safety | Bird flu |
| Random People Contact | 2.6% | 8.3% |
| Popular People Contact | 3.1% | 8.5% |
| Our Prediction Approach | **13.3%** | **19.7%** |

**Table 5. Comparison of retweeting rates.**

| Approach | Average Retweeting Rate in Testing Set under Time Constraints | |
|---|---|---|
| | Public Safety | Bird flu |
| Random People Contact | 2.2% | 6.5% |
| Popular People Contact | 2.7% | 6.4% |
| Our Prediction Approach | 13.3% | 13.6% |
| Our Prediction Approach + Wait-Time Model | **19.3%** | **14.7%** |

**Table 6. Comparison of retweeting rates with time constraints.**

Table 5 shows the comparison of retweeting rates produced by our approach against the two base lines. Overall, our approach produced a significantly higher retweeting rate than both baselines. Specifically, ours increases the average retweeting rate of two baselines by 375% (13.3% vs. 2.8%) in the public safety domain, and by 135% (19.7% vs. 8.4%) in the bird flu scenario.

**Adding Wait Time Constraint.** We also tested our wait-time model that predicts when a person would retweet after receiving a request. We compared the retweeting rate obtained using our approach with the wait-time model with that of three settings: (a) random user contact, (b) popular user contact, and (c) our approach without the use of the wait time model. In this experiment, the retweeting rate was the ratio of the people who retweeted our messages *within* the allotted time and the total number of people whom we contacted. In other words, if a person retweeted a requested message after the allotted time (e.g., 24 hours), s/he would be considered a non-retweeter as s/he did not meet the time constraint.

In our approach with the wait-time model, we set the cut-off probability at 0.7. As described previously, we first selected a subset of people who were predicted as retweeters and then eliminated those whose estimated probability to retweet within the given time window was smaller than the cut-off probability. We experimented with different time windows, such as 6, 12, 18 or 24 hours. Table 6 shows our experimental results with the averaged retweeting rates obtained for both of our data sets. Overall, our approach with the wait-time model outperformed the other three approaches in both data sets, achieving a 19.3% and 14.7% retweeting rate, respectively. Specifically, our model with wait time constraint increases the average retweeting rate of two baselines by 680% (19.3% vs. 2.45%) in the public safety domain, and by 130% (14.7% vs. 6.45%) in the bird flu scenario. This is also an improvement of 45% (19.3% vs. 13.3%) in the public safety domain and 8% (14.7% vs.

13.6%) in the bird flu domain over our own algorithm when wait time model was not used. In summary, the *combined approach* of using our prediction model and wait-time estimation further improved retweeting rates.

**Effects of Benefit and Cost.** As described previously, another method of evaluating the performance of our work is via a benefit-cost analysis using the notion of information reach. We compared the results obtained during data collection with the results of our best prediction results on the testing set. Table 7 shows the comparison of random user contact, popular user contact, and our approach without or with the wait-time model. The results show that our approach with/without the wait-time model achieved higher unit-info-reach per person than the two baselines. In particular, our approach with the wait-time model increased the average information reach of two baselines by 1,700% (153 vs. 8.5 = avg (6, 11)) in public safety and 54% (155 vs. 100.5 = avg (85, 116)) in bird flu case, respectively.

**Live Experiments**

To validate the effectiveness of our approach in a *live* setting, we used our recommender system to test our approach against the two baselines (random people contact and popular people contact). First, we randomly selected 426 candidates who had recently tweeted about "bird flu" during July 2013. We then used each approach to select 100 users among the candidates. The popular people contact and our approach selected the top 100 candidates based on their popularity (number of followers) rank and our prediction rank, respectively. If a person happened to be selected by more than one approaches, we contacted the person only once to avoid overburdening the person. Overall, we contacted a total of 232 unique people. Table 8 shows the comparison of retweeting rates for each approach. Our approach outperformed two baselines in a live setting significantly. Specifically, it increases the average retweeting rate of two baselines by more than 190% (19% vs. 6.5%). We checked the social graph of the retweeters (those who retweeted our message). They were not connected at all. Thus, our result was unlikely to be affected by their social relationship.

We also wanted to investigate the effectiveness of our approach with time constraints. Thus, we repeated the above experiment with different time windows, such as 6, 12, 18 or 24 hours. Table 9 shows the comparison of retweeting rates for each approach. Again, our approach with our wait time model outperformed all other three approaches. It in-creases the average retweeting rate of two baselines by more than 190% (18.5% vs. 6.35%). This is also an improvement of 3% over our own algorithm when the wait time model was not used. In summary, this result confirms that our approach consistently outperformed others in a live setting by a large margin.

| Approach | Unit-Info-Reach-Per-Person | |
|---|---|---|
| | Public Safety | Bird flu |
| Random People Contact | 6 | 85 |
| Popular People Contact | 11 | 116 |
| Our Prediction Approach | 106 | 135 |
| Our Prediction Approach + Wait-Time Model | **153** | **155** |

**Table 7. Comparison of information reach.**

| Approach | Retweeting Rate |
|---|---|
| Random People Contact | 4% |
| Popular People Contact | 9% |
| Our Prediction Approach | 19% |

Table 8. Comparison of retweeting rates in live experiment.

| Approach | Average Retweeting Rate |
|---|---|
| Random People Contact | 4% |
| Popular People Contact | 8.7% |
| Our Prediction Approach | 18% |
| Our Prediction Approach + Wait time model | 18.5% |

Table 9. Comparison of retweeting rates in live experiment (with time constraints).

## DISCUSSION
Here we discuss several of observations during our investigation and the limitations of our current work.

### Why People Retweet at a Stranger's Request
Although previous studies discuss various reasons why people retweet in general [4, 28], they focus on people's voluntary retweeting behavior. We were curious to find out why people retweet upon the request of a stranger. We randomly selected 50 people who retweeted per our request and asked them why they chose to retweet. 33 out of 50 replied to us. Their responses revealed several reasons why people accept our retweeting requests. One reason was the trustworthiness of the content to be spread: *"Because it contained a link to a significant report from a reputable media news source"*. Another reason is content relevance, e.g., messages about their own local area: *"Because it happened in my neighborhood"*. Interestingly, several mentioned that they retweeted because the message contained valuable information and was helpful to society: *"my followers should know this or they may think this info is valuable"*. Some of other reasons, such as to spread tweets to new audience or to entertain a specific audience, were discussed by others [4], however not mentioned in our context. In future, it would be interesting to study whether including the rationale in a retweeting request would help motivate the target strangers and affect the retweeting rate.

### Retweeting with Modification
We have observed that some people retweeted our messages with modifications (e.g., adding hashtags to clarify the message or their own opinion to the original message):

*#publichealth news: The Evolution of Bird Flu, and the Race to Keep Up http://nyti.ms/Qf6zsM @nytimesscience*

*what a shame + waste of tax $$ "@BayPublicSafety: @esavestheworld "Hacker created fake Sierra LaMar posting http://bit.ly/Leaojo" Plz RT"*

Such behavior suggests that the target information propagators may augment/alter the original message with additional information including their personal opinions, especially if they strongly agree/disagree with the intended information. Based on this observation, it would be interesting to investigate the additional gains and risks that a potential information propagator might bring when asked to spread the message. For example, the added hashtag (#publichealth) in the re-tweet above would help propagate the message not only to the followers but also those who follow the hashtag. On the opposite, a propagator's negative opinions may affect the spread and perception of the intended message.

### Generalizability
We wanted to examine how well our findings can be generalized across topics. We ran an experiment where we combined the training and test sets of public safety and bird flu. We trained prediction models on the combined training set using the significant features identified for the combined set. AUC in this experiment was 0.736, better than the original public safety result (0.692), but lower than the original bird flu result (0.785). The resulted retweeting rate was 12.5%, better than the random user contact (5.5%) and popular user contact (6%) for the combined set, but lower than the rates achieved in public safety (13.3%) and bird flu alone (19.7%). Our results suggest that it is feasible to build a domain-independent prediction model, if we have sufficient training-data from different domains. We are investigating the applicability of our models to new domains, e.g., new topics that our model is not trained on.

### Optimizing Multiple Information Spreading Objectives
Currently, our work focuses on maximizing the retweeting rate in information diffusion. However, in practice, there may be multiple objectives to be satisfied, such as maximizing the expected net benefit or minimizing the reach time. We thus are investigating a model that can optimize multiple objectives at the same time. However, this is non-trivial as satisfying one objective may influence the other especially in a real world situation, where many of these objectives may be dynamically changing (e.g., the availability of retweeting candidates and the required time frame for a message to reach a certain audience).

## CONCLUSIONS
In this paper, we have presented a feature-based prediction model that can automatically identify the right individuals at the right time on Twitter who are likely to help propagate messages per a stranger's request. We have also described a time estimation model that predicts the probability of a person to retweet the requested message within a given time window. Based on these two models, we build an interactive retweeter recommender system that allows a user to identify and engage strangers on Twitter who are most likely to help spread a message. To train and test our approach-es, we collected two ground-truth datasets by *actively* engaging 3761 people on Twitter on two topics: public safety and bird flu. Through an extensive set of experiments, we found that our approaches were able to at least *double* the retweeting rates over two baselines. With our time estimation model, our approach also outperformed other approaches significantly by achieving a much higher retweeting rate within a given time window. Furthermore, our approach has achieved a higher unit-information-reach per person than the baselines. In a live setting, our approach consistently outperformed the two baselines by almost doubling their retweeting rates. Overall, our approach effectively identifies qualified candidates for retweeting a message within a given time window.

**REFERENCES**

1. Agarwal, N., Liu, H., Tang, L., and Yu, P. S. Identifying the influential bloggers in a community. In.*WSDM*, 2008.
2. Bakshy, E., Hofman, J. M., Mason, W. A., and Watts, D. J. Everyone's an influencer: quantifying influence on twitter, In *WSDM,* 2011.
3. Bakshy, E., Rosenn, I., Marlow, C., and Adamic, L. The role of social network in information diffusion. In *WWW,* 2012.
4. Boyd, D., Golder, S., and Lotan, G. Tweet, Tweet, Retweet: Conversational Aspects of Retweeting on Twitter. In *HICSS,* 2010.
5. Budak, C., Agrawal, D., and El Abbadi, A. Limiting the spread of misinformation in social networks. In *WWW*, 2011.
6. Cha, M., Haddadi, H., Benevenuto, F., and Gummadi, K.P. Measuring user influence in twitter: The million follower fallacy. In *ICWSM*, 2010.
7. Chaoji, V., Ranu, S., Rastogi, R., and Bhatt, R. Recommendations to boost content spread in social networks., In *WWW*, 2012.
8. Chawla, N. V., Bowyer, K. W., Hall, L. O., and Kegelmeyer, W. P. SMOTE: Synthetic Minority Over-sampling Technique. *Journal of Artificial Intelligence Research*, 16: 321-357, 2002.
9. Chen, J. Cypher, A., Drews, C. and Nichols, J. CrowdE: Filtering Tweets for Direct Customer Engagements. In *ICWSM* 2013.
10. Costa, P.T., and McCrae, R.R. Revised NEO Personality Inventory (NEO-PI-R) and NEO Five-Factor Inventory (NEO-FFI) manual. *Psychological Assessment Resources*, 1992.
11. Fast, L. A., and Funder, D. C. Personality as manifest in word use: Correlations with self-report, acquaintance report, and behavior. *Journal of Personality and Social Psychology*, Vol 94(2), 2008.
12. Fawcett, T. An introduction to ROC analysis. *Pattern Recogn. Lett.*, Vol 27( 8), 2006.
13. Gill, A. J., Nowson, S., and Oberlander, J. What Are They Blogging About? Personality, Topic and Motivation in Blogs, In *ICWSM, 2009*.
14. Goyal, A., Bonchi, F., and Lakshmanan, L. V.S. Learning influence probabilities in social networks. In *WSDM*, 2010.
15. Hall, M., Frank, E., Holmes, G., Pfahringer, B., Reutemann, P., and Witten, I. The WEKA data mining software: an update. *SIGKDD Explorations Newsletter*, 11(1): 10-18, 2009.
16. Hoang, T.-A., and Lim, E.-P. Virality and Susceptibility in Information Diffiusions, In *ICWSM* 2012.
17. Huang, J., Cheng, X.-Q., Shen, H.-W, Zhou, T., and Jin, X. Exploring social influence via posterior effect of word-of-mouth recommendations. In *WSDM*, 2012.
18. Lee, K., Caverlee, J., and Webb, S. Uncovering social spammers: social honeypots + machine learning. In *SIGIR* 2010.
19. Lee, K., Eoff, B. D., and Caverlee, J. Seven Months with the Devils: A Long-Term Study of Content Polluters on Twitter. In *ICWSM,* 2011.
20. Liu, X.Y., and Zhou, Z.H. The influence of class imbalance on cost-sensitive learning: an empirical study. In *ICDM, 2006*.
21. Macskassy, S. A., and Michelson, M. Why Do People Retweet? Anti-Homophily Wins the Day!. In *ICWSM*, 2011
22. Mahmud, J., Zhou, M., Megiddo, N., Nichols, J., and Drews, C. Recommending Targeted Strangers from Whom to Solicit Information in Twitter. In *IUI,* 2013.
23. Nichols, J., and Kang, J-H. Asking Questions of Targeted Strangers on Social Networks. In *CSCW,* 2012.
24. Pennebaker, J.W., Francis, M.E., and Booth, R.J. Linguistic Inquiry and Word Count. *Erlbaum Publishers*, 2001.
25. Romero. D. M., Galuba. W., Asur. S, and Huberman, B. A. Influence and passivity in social media. In *ECML/PKDD,* 2011.
26. Shannon, C. E., A mathematical theory of communication. Bell system technical journal, Vol 27, 1948.
27. Singer, Y. How to win friends and influence people, truthfully: influence maximization mechanisms for social networks, In *WSDM*, 2012.
28. Starbird, K. and Palen, L. Pass It On?: Retweeting in Mass Emergency, In *ISCRAM*, 2010.
29. Ver Steeg, G. and Galstyan, A. Information transfer in social media. In *WWW*, 2012.
30. Weng, J. Lim. E.-P., Jiang. J, and He. Q. Twitterrank: Finding topic-sensitive influential twitterers. In *WSDM*, 2010.
31. Yang, Y., and Pedersen, O.J. A Comparative Study on Feature Selection in Text Categorization. In *ICML*, 1997.
32. Yarkoni, Tal. Personality in 100,000 words: A large-scale analysis of personality and word usage among bloggers. *Journal of Research in Personality,* 2010.